# A simple strategy for enhanced production of nanoparticles by laser ablation in liquids

**Yaakov Monsa[1], Gyora Gal[2], Nadav Lerner[3] and Ilana Bar[1]**

[1] Department of Physics, Ben-Gurion University of the Negev, Beer-Sheva 8410501, Israel.
E-mail address:
[2] Department of Chemistry, Nuclear Research Center Negev, P.O. Box 9001, Beer-Sheva 8419001, Israel.
[3] Department of Analytical Chemistry, Nuclear Research Center Negev, P.O. Box 9001, Beer-Sheva 8419001, Israel

E-mail: ibar@bgu.ac.il

## Abstract

Upgrading the productivity of nanoparticles (NPs), generated by pulsed laser ablation in liquid (PLAL), still remains challenging. Here a novel variant of PLAL was developed, where a doubled frequency Nd:YAG laser beam (532 nm, ~ 5 ns, 10 Hz) at different fluences and for different times was directed into a sealed vessel, toward the interface of the meniscus of ethanol w ith a tilted bulk metal target. Palladium, copper and silver NPs, synthesized in the performed proof of concept experiments, were mass quantified, by an inductively coupled plasma optical emission spectrometry, and characterized by ultraviolet-visible extinction spectroscopy, transmission electron microscopy and X-ray diffraction. The NPs consist of crystalline metals of a few nm size and their ablation rates and agglomeration levels depend on the employed laser fluences. The ensuing laser power-specific productivity curves for each metal, peaked at specific laser fluences, were fitted to the results of a simple model accounting for plasma absorption and heat transfer. The resulting peaked yields and concentrations were more than an order of magnitude higher than those obtained for totally immersed targets. Besides, the measured productivities showed nearly linear dependencies during time intervals up to 30 min of ablation, but became saturated at 1 h, due to accumulation of a significant number of NPs along the laser beam path, reducing the laser intensity reaching the target. This suggested approach could inspire future studies that will contribute to further developments of efficient generation of NPs with controlled characteristics.

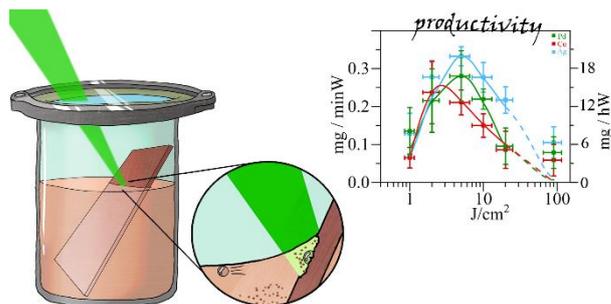

Supplementary material for this article is available in the end of this manuscript

## 1. Introduction

Laser ablation of solid materials attracts considerable interest and is in the focus of extensive research [1-4], due to its great potential for laser based materials processing, involving thin solid film preparation, nanoparticles (NPs) synthesis, surface cleaning, and laser cutting, welding and drilling. Usually, it is performed in deposition chambers [1, 2, 5] under vacuum, or filled with gas, and otherwise in vessels containing targets immersed in various liquids [3, 4]. The latter method, namely, pulsed laser ablation in liquid (PLAL), became recently a promising simple approach for synthesis of nanomaterials and nanostructures of high purity, requiring just irradiation of a solid target immersed in a liquid by short, or ultrashort pulsed laser beams [3, 4, 6-13].

The key steps of PLAL [6, 7, 9, 14], include laser pulse absorption by the bulk target, resulting in its initial heating and subsequent plasma plume generation and expansion into the surrounding liquid, leading to a shockwave. During the expansion and condensation, the plasma plume cools down, through energy release to the liquid, to generate a cavitation bubble that expands in the liquid, till its internal pressure is lost. Then, the cavitation bubble collapses by the liquid pressure back to its origin and the collapse energy is translated back to a shockwave, heating the liquid and the solid within the ablation crater, creating an additional bubble and so on till full energy dissipation occurs. Upon plasma cooling, NPs are formed and dispersed into the surrounding liquid, during bubble collapse, to form a colloidal solution.

In spite of the complex cascade of physical-chemical phenomena, taking place during PLAL, it offers substantial advantages. These advantages include a fairly simple experimental setup with the possibility to perform experiments under ambient conditions, using simple starting materials embedded in minimal amounts of nonpolluting solvents and frequently with no surfactants, or catalysts. Hence, this approach eventually leads to formation of safe and high purity products, allowing formation of NPs with compositions, morphologies, and other properties that can be controlled by varying liquids and laser parameters. In addition, the quite extreme conditions (temperatures of thousands of K and pressures of GPa) induced by PLAL, offer the possibility to create novel materials. For these reasons, PLAL is established as a continuously growing field for preparation of a large variety of metal and non-metal nanomaterials [3, 4, 6-13]. Yet, PLAL still faces major challenges, including the understanding of the fundamentals of NPs formation, modelling of the process and its outcome, control of size, shape and agglomeration of the generated NPs and the prospect of overcoming the poor productivity [6, 7, 15-18].

Commonly, PLAL is performed by focusing the laser beam on a totally immersed bulk target, being parallel to the liquid surface, leading to yields, shapes and size distributions of NPs, strongly depending on the laser wavelength, pulse duration, fluence, repetition rate and energy, as well as liquid thickness above the bulk. Essentially, the laser wavelength affects PLAL in several ways, depending on the target material and the generated NPs. For instance, for bulk metals the absorption coefficient increases with decreasing laser wavelengths [19]. However, since the laser beam is reabsorbed by the produced NPs, the ablation efficiency does not necessarily increase upon short wavelength irradiation [20]. The reabsorption stems from valence-to-conduction (interband) transitions at short ultraviolet-visible (UV-Vis) wavelengths and surface plasmon resonances (SPRs) [21, 22], i.e., collective oscillations of conduction free electrons excited by the electromagnetic field of light at longer wavelengths. Actually, NPs reabsorption can decrease the ablation rate and trigger further modifications during PLAL, decreasing their average size and broadening the size distribution [7, 23-28], or leading to submicrometer spherical particles growth [29]. Efficient absorption of laser pulses at short wavelengths can occur also by the plasma plume, especially in ns PLAL [30], introducing nonlinearity between ablation efficiency and laser fluence.

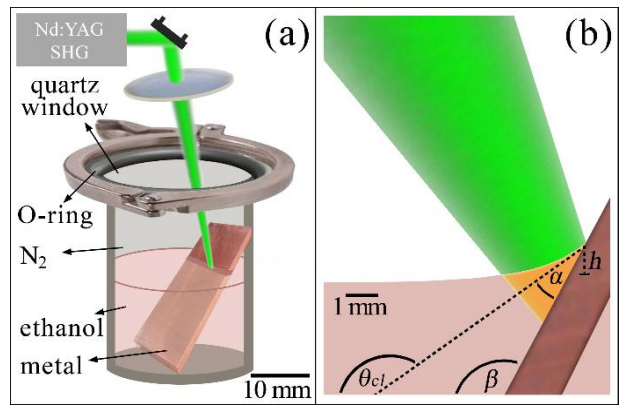

Fig. 1. (a) Schematic diagram of the experimental setup for ablation at the interface of the meniscus of ethanol with a tilted metal target. The system includes a Nd:YAG laser with a second harmonic generator (SHG), delivering a 532 nm laser beam, focused by a planoconvex lens onto the target. The beam passes through a quartz window, positioned above a Viton O-ring and fixed by a clamp, toward the target located in the sealed stainless steel vessel containing 5 ml of ethanol, under nitrogen gas at atmospheric pressure and (b) enlarged view of the largest spot size of the focused beam at the interface of ethanol-metal, sampling maximal thickness of about 2 mm of ethanol above the target, for the lowest used fluence of 1 J/cm². The meniscus shape was calculated according to Ref. [42], considering the $\alpha$, $\beta$, $\theta_{cl}$ and $h$ parameters.

These factors, as well as the laser beam interaction with the high concentration of the ablated matter close to the crater and the laser light scattering by the cavitation bubble, which reduces the laser energy reaching the target, affect extensively the NPs productivity in PLAL [31]. The former can be reduced by inclining the target with respect to the beam direction [32], while the latter by decreasing the laser repetition rate, namely the time interval between successive laser pulses ($\Delta t_P$). Actually, a linear productivity increase is expected for $\Delta t_P$ longer than the cavitation bubbles lifetime (of the order of $10^{-4}$ s) [15, 16, 33].

The relationship between productivity and laser, target and liquid parameters is still not completely known, limiting the generated yields to several, or dozens of mg/hW and hindering the development of PLAL for industrial applications.

Actually, the general approach where PLAL is performed with the laser beam focused on a totally immersed bulk target leads to poor productivities, as very recently encountered, in our production of monometallic, copper (Cu) and palladium (Pd) NPs [34]. To overcome this obstacle, different approaches were suggested and tested. For instance, they include sample movement [16], change of the target geometry to a wire [17, 18], liquid flow [16, 35, 36], application of a high-speed rotating target [37], use of a complicated system containing high power and repetition rate laser pulses with a polygon scanner operating at high speed to overcome bubble shielding [15, 36] and decrease of the liquid layer thickness [16, 38-41]. As a matter of fact, the latter factor and the target position are key parameters in controlling the size and yield of the prepared NPs in PLAL.

Here, a new, relatively simple, yet efficient approach, is suggested and investigated, using a frequency doubled, pulsed nanosecond (ns) Nd:YAG laser (532 nm, 10 Hz, ~ 5 ns) at different fluences for ablation of Pd, Cu and silver (Ag) targets, positioned at a unique geometry. Specifically, the laser beam was directed into a sealed vessel toward the interface of a meniscus formed by the liquid (ethanol) attached to a tilted bulk metal target. This geometry provided a very thin liquid layer in the irradiation region of the target and allowed generation of NPs, falling toward the vessel bottom, thus minimizing beam absorption and scattering by the emerging NPs. This approach resulted in enhanced NPs yields, depending on laser fluence, which could be interpreted in terms of plasma absorption and heat transfer.

## 2. Experimental

### 2.1 Generation of monometal nanoparticles

Colloidal metal NPs were prepared by the second harmonic (532 nm) of a pulsed Nd:YAG laser (Continuum, Surelite SLI-10) with ~ 5 ns pulse duration, operating at 10 Hz repetition rate. The bulk metal targets Pd, Cu or Ag, purchased from Aldrich, with 99.9, 99.999 and 99.9 % purities, respectively, were cut into pieces of about $25 \times 19 \times 1$ mm$^3$ and cleaned in ethanol. After drying the samples, they were introduced in a stainless steel vessel and held against its back wall [figure 1(a)], leading to a tilted metal target, partially immersed within 5 ml of absolute ethanol (Bio-Lab). The angle between the tilted target and the liquid surface ($\beta$) [Fig. 1(b)] was essentially defined by the target length. To prevent ignition, by the reaction of oxygen (available in air) with the ethanol vapor, the air above the ethanol was forced out by flushing with nitrogen at atmospheric pressure, to eventually maintain an inert atmosphere in the vessel. Then, the vessel was immediately closed by a clamp that fixed a quartz window above a Viton O-ring. The laser beam was directed and focused using various focal length quartz lenses, matching the lens-target distances, just below the ethanol/nitrogen boundary onto the interface of the tilted bulk metal target with the meniscus formed by ethanol [figure 1(b)]. This point was identified through observation of a much weaker plasma emission, simply by sliding the shinning laser beam from the bare target toward the meniscus.

Upon use of a pulse laser energy of 68 mJ on the target and the spot areas on the surface, controlled by the lens-target distances, laser fluences in the 1-90 J/cm$^2$ range were obtained. The laser energies were measured by a thermopile power meter (Coherent, FieldMate with a PM10 sensor) and the laser spot sizes were measured by taking single shot burns on a photosensitive paper (Kodak, Linagraph 1895) positioned on the target. The measured burns matched the calculated beam sizes for a Gaussian beam propagating in free space and following focusing with the different focal length lenses. The different targets were initially irradiated for 5 min, i.e., 3000 laser pulses at 10 Hz repetition rate, to generate the corresponding colloids. For comparison, also top ablation experiments, of 5 min duration with the incident laser beam focused on parallel targets positioned at the vessel bottom and embedded into 5 ml of ethanol (~ 4 mm height) were performed. In an additional set of experiments the targets were irradiated for different time intervals, up to 1 h.

### 2.2 Nanoparticles characterization

The produced NPs were examined by several means. In particular, their mass was quantified by employing an inductively coupled plasma optical emission spectrometer (ICP-OES, SPECTRO model ARCOS, side on plasma configuration) for measurement of the concentrations of Pd$^{2+}$, Cu$^{2+}$ and Ag$^+$ ions, with an error limit of ± 10 %. Three sets of Pd, Cu or Ag samples, with a known amount of their colloidal suspensions in ethanol, were dried by heating and then dissolved using 6 ml of hot piranha solution (sulfuric acid and hydrogen peroxide, 5:1 ratio). Each sample was diluted using ultrapure water (milli-Q setup, Millipore, 18.2 MΩ) to a final volume of 0.5 l. Sample concentrations were measured using four-point calibration curves (concentrations: 0, 0.5, 2.5, 5 mg/l). The amount of Pd, Cu, or Ag in the colloidal suspensions was calculated considering the final dilution factor. All the calibration solutions were diluted from 1 g/l standard solutions, purchased from CPA. These measurements allowed to determine the productivity curves for each metal. In addition, the ablation mass of the NPs dispersed in ethanol was deduced for each experiment by weighing the target before the ablation process and after drying the target following the ablation by a digital analytical balance (BOECO Germany, BAS32) with 0.1 mg readability.

The UV-Vis extinction spectra of the colloidal NPs were measured through quartz cells (optical path lengths of 10 mm)

on a spectrophotometer (Thermo Scientific, Genesys 10S) in the 200 - 800 nm range. These spectra of the samples were measured, using pure ethanol as blank, following dilution of 1 ml of each suspension in 2 ml of ethanol, to avoid detector saturation.

The morphologies and size distributions of the particles were characterized using a transmission electron microscope (TEM) (FEI, Tecnai 12 G$^2$ TWIN). The TEM samples were prepared by dripping colloids of each of them on an ultrathin carbon film supported by a lacey carbon film on a 300 mesh gold, or a 400 mesh copper grid (Ted Pella Inc.). Sample characterization was performed by the XRD method for revealing the crystal structure of the NPs. A powder diffractometer (PANalytical Empyrean, Almelo) equipped with position sensitive detector X'Celerator was used. Data were collected in the θ/2θ geometry using Cu Kα radiation (λ=1.54178 Å) at 40 kV and 30 mA. Scans were run during ~15 min in a 2θ range of 10 - 80º with steps equal to ~ 0.033º.

## 3. Results and Discussion

### 3.1 Concept and design

The Pd, Cu and Ag NPs were generated following the use of the system shown in figure 1(a), where the laser beam was focused at the meniscus interface, created by ethanol with the inclined metal target, figure 1(b). This unique system was devised, foreseeing that it could provide thin liquid layers in the ablation region, which would possibly affect ablation productivities (see below).

The meniscus shape of the liquid, attached to the inclined plate [figure 1(b)] was computed by implementing the numerical method of Pozrikidis *et al*. [42] in a Mathematica code, using the NDSolve function. The numerical calculation was based on the solution of a first-order ordinary differential equation [42]:

$$\frac{df}{dx} = -\left[\left(\frac{2}{2-\frac{f^2}{l^2}}\right)^2 - 1\right]^{1/2}, \qquad (1)$$

where $f$ is the meniscus elevation as a function of the distance $x$ from the point of attachment of the liquid-target plate and $l \equiv \left(\frac{\gamma}{\rho g}\right)^{1/2}$ is the capillary length, with $\gamma$, $\rho$ and $g = 9.8$ m/s$^2$ corresponding to the liquid surface tension, the liquid density and the acceleration of gravity, respectively. The boundary condition for the elevation of the interface at the contact line [figure 1(b)] is given by:

$$h = \sqrt{2}l(1 - |cos\theta_{cl}|)^{\frac{1}{2}} \qquad , \qquad (2)$$

where $\theta_{cl} = \alpha + \beta$, corresponding to the $\frac{1}{2}\pi < \theta_{cl} < \pi$ range. Actually, $\alpha$ is related to the contact angle of ethanol with the metals and $\beta$ to the inclination angle of the target. Considering $\alpha$ for outflow contact angles of ethanol with Cu and Ag metals, 0.12 rad [43], respectively, $\gamma = 21.78$ mN/m (at 20 ºC) and $\rho = 790.0$ kg/m$^3$ values for ethanol [44], the resulting meniscus shape could be obtained. This shape actually determines the accessed liquid layer thickness above the target, upon positioning the incident laser beam, of different spot sizes for various fluences, just below the meniscus edge. Based on the above, the menisci shapes for other liquids should be slightly different, see figure S1 of the Supplementary Material. Nevertheless, the differences for ethanol, water and acetone are quite small, implying that the productivities, in this aspect, will not be affected too much.

As can be seen from figure 1(b), the maximal calculated thickness of the layer of ethanol above the target, for a target inclined at $\beta = 2.1 \pm 0.1$ rad was about $2.0 \pm 0.2$ mm for the spot size, corresponding to 1 J/cm$^2$ laser fluence. This thickness was even lower for the higher used fluence beams, i.e., $0.7 \pm 0.2$ mm for 20 J/cm$^2$. This target inclination angle ($\beta = 2.1 \pm 0.1$ rad) was selected since it provides low thicknesses of ethanol above the targets, but yet deviation of the incident beam from the bubbles and the NPs, see figure S2 of the Supplementary Material. Actually, this angle tolerates ethanol levels above the targets, for higher fluences, in the preferable 1 mm regime [16, 38-41], probably avoiding too thin layers. This is of importance for precluding the formation of part of the plasma plume outside the liquid [45], which can occur for liquid layer thicknesses approximating the laser-induced plasma size. It is obvious that the experiments can be performed at additional inclination angles of the targets and yet remains to be tested whether this angle is optimal for obtaining the highest productivities.

As mentioned above, fluence control was achieved by exploiting a fixed laser energy and planoconvex lenses with various focal lengths, matching the lens-target distances, which resulted in smaller beam diameters and somewhat lower thicknesses for higher fluence laser beams. This mode was chosen to assure measurable yields in short ablation times. Another option would of course be to induce ablation by alterable laser energies, while keeping the focal length constant. This approach would be preferable, since it excludes effects of focal spot area and liquid thickness on the productivity, however, in this type of experiments the laser intensity has to be reduced by more than an order of magnitude to obtain the low fluences, which would lead to barely measurable productivities during the used ablation times. Moreover, it is conceivable that although the range of employed liquid thicknesses is nonuniform, the ablation volumes should be comparable (within error) for a Q-switched Nd:YAG laser [38]. Recalling this behavior and the slight changes in liquid thicknesses in our experiment, the change of focal lengths was chosen to provide the tested fluences.

By directing the laser beam toward the inclined metal targets, through the low thicknesses ethanol layers, small NP concentrations in the irradiated regions were expected. This specific configuration was considered to have the potential to overcome the lower liquid thickness limit set by plasma confinement and bubbles onset, which break through the liquid-air interface and lead to splashing [6]. Actually, we have noticed that the pressure in the vessel raised, during the ablation, leading to minimized splashing. This is a consequence of the induced evaporation of the liquid, degassing and liquid splitting, by the high temperature of the laser-induced plasma [46, 47].

Considering that usually the plasma plume, ablated particles and cavitation bubbles are positioned along the laser beam and perpendicular to the target surface [6, 7, 9], while here they are inherently spatially bypassed through reirradiation by successive pulses, it is supposed that NPs absorption is prevented, probably affecting the ablation productivity. At higher repetition rates (1 kHz and above)] [32], the shielding and reflection of the incoming pulse by cavitation bubbles and consequently the reduction of the pulse energy that reaches the target surface should be also significant. Accordingly, it is believed that the conceived configuration might play a significant role in PLAL experiments at higher repetition rates, providing bubbles propagation, deviating from the laser beam path.

### 3.2 Nanoparticles productivities

Preliminary experiments, already confirmed that the dispersions of (a) Pd, (b) Cu and (c) Ag NPs, produced during 5 min ablation at the interface of the meniscus of the metal targets with ethanol [$\beta = 2.1 \pm 0.1$ rad, see figure 1(b)], shown in figure 2(A), were very dark. This is particularly so if they are compared to the respective dispersions obtained by top ablation, figure 2(B). The considerably darker suspensions obtained by inclined targets ablation indicate that the productivity in ablation at the interface of the meniscus is substantially higher.

Laser power-specific NP productivities were obtained by accounting for the ablated mass in each sample measured by ICP-OES, the laser fluence used for ablation and the 5 min process duration. Figure 3 presents the power-specific productivities obtained under these conditions, for (a) Pd, (b) Cu and (c) Ag NPs, as a function of the employed laser fluences. It can be clearly seen that the curves, peaked at specific productivities, achieved at certain laser fluences, by ablation at the interface of the meniscus of ethanol and the tilted targets. This dependence of productivity on fluence is in agreement with previous studies on PLAL [17, 48, 49] and ablation in vacuum [50], although in the former only very low fluences were examined [17]. It is interesting to note, that the masses deduced by ICP-OES agreed, within experimental error, with those retrieved by weighing, implying that the generated NPs are captured by the liquid and do not move into the nitrogen, similar to previous findings [46, 47].

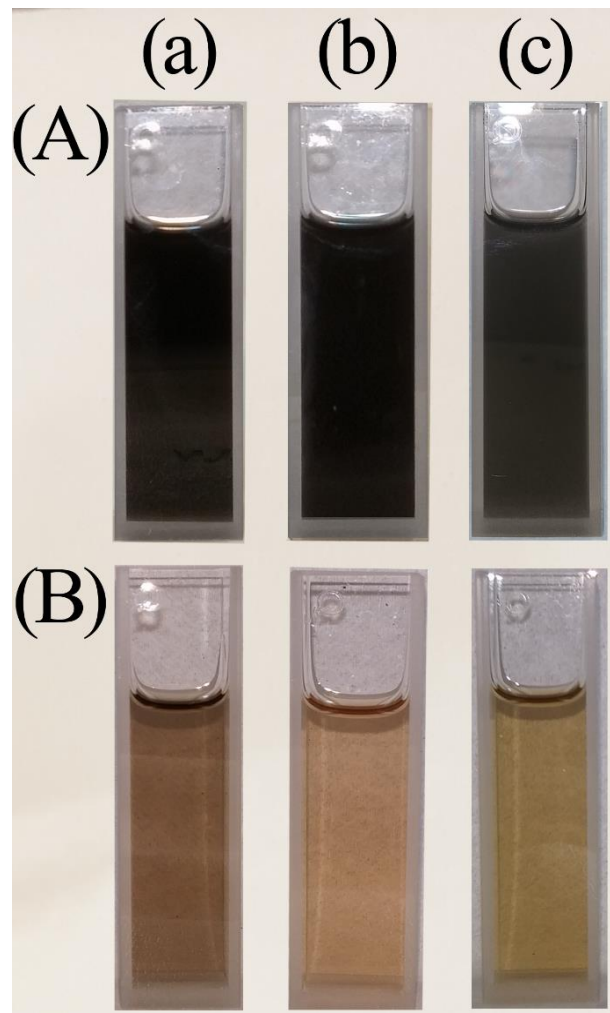

Fig. 2. Ethanol dispersions in cuvettes of (a) Pd, (b) Cu and (c) Ag nanoparticles, obtained following 5 min of ablation. The dispersions exhibit those obtained by ablation at the interface of the meniscus of the corresponding metal targets with ethanol, $\beta = 2.1 \pm 0.1$ rad, in (A) and those by top ablation of the targets, embedded in 5 ml ethanol (~ 4 mm height), in (B).

Generally, the curve shape of the dependence of ablation productivity and the ablation depth of a variety of targets, on laser fluence was elucidated previously and even treated by some models [30, 51-53], assuming that ns laser ablation is a thermodynamic process consisting of several steps. Specifically, in the low fluence regime, the very slow heating rate of the target (related to its thermal conductivity) leads to inefficient ablation since the surface could not reach appropriate high temperatures. For laser fluences that can achieve critical temperatures, i.e., threshold fluences for ablation, a sharp increase in ablation depth could occur due to phase explosion. This leads to significant decrease in absorption coefficients and thermal conductivities of the targets, as well as in their reflectivity at the dielectric-metal interface [52]. At this point the ablation productivities start to grow by rising the laser fluences, however, too high laser fluences can lead to matter ionization, so that the intense formed plasma absorbs the incoming laser beam, reducing the productivity [30, 52]. Hence, this behavior causes growth and subsequent decay of the ablation depth for higher laser fluences, reaching maximum productivity at a specific fluence, supporting the general trend shown by the productivity curves in figure 3.

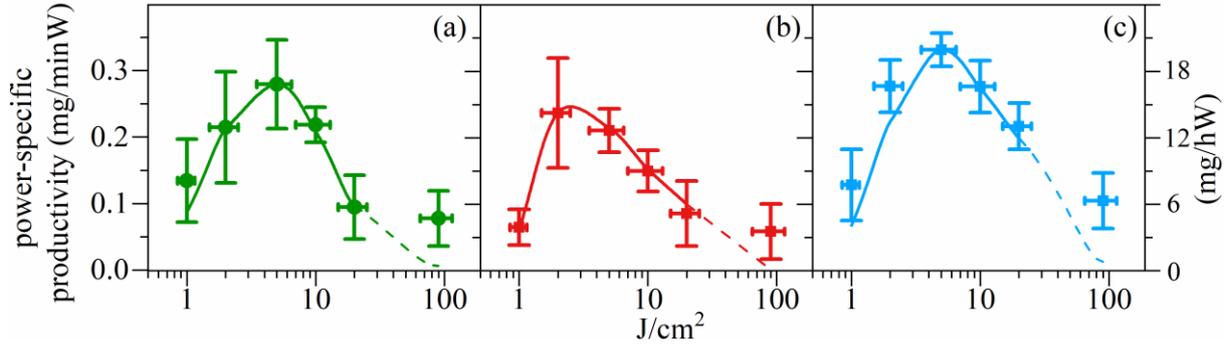

Fig. 3. Measured (scattered) and calculated (lines) power-specific productivities (mg/minW with extrapolation to mg/hW) for (a) palladium, (b) copper and (c) silver as a function of laser beam fluences. The calculated curves were obtained by fitting the productivities in the range of low (solid lines) effective laser fluences, $F_{eff}$, for specific simulated plasma absorption coefficients and continued to the high (dashed lines) $F_{eff}$.

A quantitative analysis of the resulting curves is challenging, however, in terms of the above features, a Matlab code was written, in attempt to fit the laser power-specific NPs productivities for Pd, Cu and Ag *vs.* the laser fluences, shown in figures. 3(a), 3(b) and 3(c), respectively. It was assumed that the ablated matter should be proportional to the incident laser energy. However, since during the laser pulse the expansion of the laser-induced high density plasma is strongly confined by the liquid [6, 7, 9], energy losses by absorption occur [30, 52]. These losses are also accompanied by heat transport into the target [51, 52], implying that only an "effective" laser fluence, $F_{eff}$, could be involved in the ablation.

Therefore, we aimed to estimate $F_{eff}$, influenced by the above mentioned factors, plasma absorption and heat transport. The former is proportional to an exponent, depending on the absorption coefficient, $\alpha_p$, and the plasma thickness, or ablation depth, $d$ [30, 52]. The latter is related to the ablated volume-to-surface ratio, $\frac{V_a}{S_a}$, where larger surface areas correspond to faster heat transfer and hence lower $F_{eff}$. Since the laser beam waist radius is typically at least three orders of magnitude larger than the optical absorption depth in metals [52], it is considered that $\frac{V_a}{S_a} \approx d$ (estimated from the measured ablated mass and the beam spot size). Consequently, the $F_{eff}$ that induces ablation is proportional to:

$$F_{eff} \propto e^{-\alpha_p d} d \quad . \qquad (3)$$

This implies that for fluences in the regime below and above the peak productivities, figure 3, the term related to plasma absorption in Eq. (3) is less and more dominant, respectively, while the heat transfer term promotes a contrary behavior. By fitting the results of this model to the measured productivities in the low fluence regime, $\alpha_p$ values of $7 \times 10^7$, $8 \times 10^7$ and $4 \times 10^7$ m$^{-1}$ for Pd, Cu and Ag, respectively, were obtained.

As can be seen from figure 3, panels (a) – (c), these fits (lines) are quite compatible with the measured data (scattered points) in the low fluence regime (solid lines), implying that for rising laser fluences in the low range, concomitant increase of energy deposited in the plasma and decrease of that supplied to the target occurred. Nevertheless, at high fluences the measured and calculated (dashed lines) values depart, since the laser fluence in this regime is high enough to induce liquid breakdown [19, 54], or at least plasma at the air-water interface, progressively absorbing a larger portion of laser energy [48]. This simplistic model gives some insight into the process, but does not take into account other parameters, like hardness and melting point of the target [55], as well as such as thermal conductivities, heat capacities, or optical penetration depths, which also affect the ablation efficiency and might be competing parameters for this laser-target-liquid system.

The revealed maximal laser power-specific NP productivities of 0.28, 0.24 and 0.33 mg/minW, figure 3, for 5 min ablation of Pd, Cu and Ag at laser fluences of 5, 2 and 5 J/cm², respectively, might be extrapolated to 16.8, 14.4 and 19.8 mg/hW. These productivities are somewhat higher than those obtained by direct 1 h laser ablation of the different targets, in the preferable fluence regimes [figures 4(a)-(c)], namely, 11.8, 7.64 and 14.14 mg/hW for Pd, Cu and Ag targets. Although the productivities show approximately linear dependence during time intervals up to 30 min of ablation, they become saturated at 1 h. The decrease in ablation efficiency at the longest ablation time occurs because of the small vessel in which the experiment was performed, probably setting a significant number of NPs along the laser beam path, thus reducing the intensity reaching the target. This difficulty can be resolved by using a larger cell, or by removing the as-prepared NPs by a flow-through system.

Nevertheless, these productivities, obtained with the simple system and configuration suggested here, consisting of a static vessel and a laser beam directed toward the interface of the meniscus of ethanol with the tilted bulk metal target, are quite impressive. It is particularly so, if it is considered that they are more than an order of magnitude larger, than in top ablation [figure 2(B)] and in similar experiments on Pd [48]. Moreover, these productivities are quite remarkable, if compared to the high productivities for Cu (5.4 mg/hW) and Ag (6.8 mg/hW) NPs [6, 56], obtained by ablation with the fundamental of a Nd:YAG laser of a one dimensional wire in a static chamber, containing water. Evidently, it is anticipated that our productivities, could be upgraded by using a larger volume vessel, by integration of liquid flow [4, 6, 16, 35, 36], or by sample movement [4, 6, 16]. It is worth mentioning an additional advantage of these experiments, namely the extremly high concentrations that could be achieved, i.e., 1.0 - 1.9 mg/ml for ablation experiments of 60 min, figures 4(a)-(c).

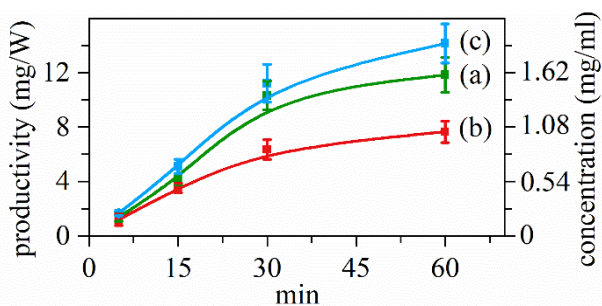

Fig. 4. Measured power-specific productivities (mg/W) and concentrations (mg/ml) for (a) palladium, (b) copper and (c) silver as a function of the time, for the preferable laser fluences of each metal, shown in Fig. 3. The plotted lines are guide to the eyes.

### 3.3 Nanoparticles characterization

The formation of Pd, Cu and Ag NPs was monitored by UV-Vis extinction spectroscopy. The resulting originally measured UV-Vis absorbance spectra of the corresponding NPs, prepared with different laser fluences are shown in figure S3 of the Supplementary Material. These spectra are similar to the vertically translated ones, shown in figures 5(a), 5(b) and 5(c), for the respective NPs, prepared by laser fluences of (*i*) 1, (*ii*) 2, (*iii*) 5, (*iv*) 10, (*v*) 20 and (*vi*) 90 J/cm². The spectra of the different colloidal dispersions in the 200 - 800 nm range, measured immediately after ablation, display different shapes. It turns out that all the spectra exhibit dominating and poorly resolved UV absorption bands centered around 210 nm with pronounced tails extending down to 800 nm, while those of Cu and Ag NPs are significantly disturbed by the additional bands centered near 580 and 408 nm, respectively, attributed to their SPRs [21, 22]. Although slight changes appear in the extinction spectra of the Pd, Cu and the Ag NPs, obtained by ablation with different laser beam fluences, the most prominent emerge while passing from 10 to 20 J/cm². Consequently, these spectra are considered, in attempt to enlighten the dynamics of the NPs generation.

In particular, for Pd NPs, the spectra [figure 5(a)] are characterized by low wavelength peaks with decaying absorption intensities toward higher wavelengths, where in (*iv*) for 10 J/cm² the drop in absorption is substantially higher than that in (*v*) for 20 J/cm². As for Cu NPs [Fig. 5(b)], a slight declining slope of the UV absorption, with a huge and broad SPR peak situated at 610 - 620 nm, is encountered in (*iv*) for

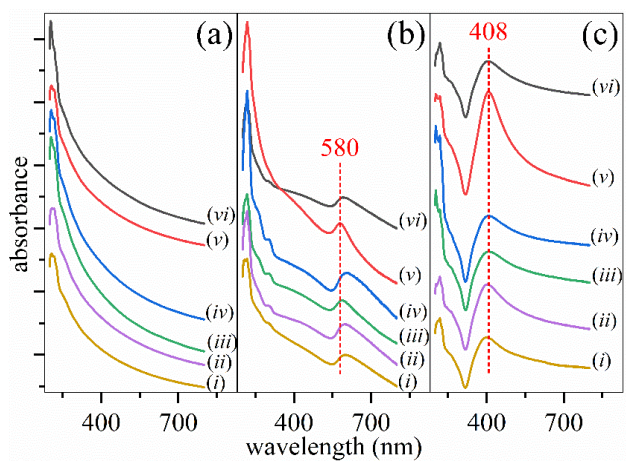

Fig. 5. UV-Vis extinction spectra of (a) Pd, (b) Cu and (c) Ag nanoparticles, ablated from the corresponding tilted metal targets, at the interface with the meniscus of ethanol, using beam fluences of (*i*) 1, (*ii*) 2, (*iii*) 5, (*iv*) 10, (*v*) 20 and (*vi*) 90 J/cm². The wavelengths and the dashed lines mark peak positions of the surface plasmon resonances in (b) and (c). The dashed lines emphasize the relative positions of the peaks in the samples prepared under different fluences.

10 J/cm². This contrasts the most intense slope of extinction in (*v*) for 20 J/cm², associated with th e most outstanding and narrowest SPR peak, positioned at 580 nm and considered as characteristic for Cu NPs [57]. In addition, the growth of two peaks around 260 and 300 nm is encountered for the low laser fluences, i.e., (*i*) 1 to (*iv*) 10 J/cm², but not for (*v*) 20 J/cm². These two peaks resemble those seen previously [24, 58-59], which were suggested to be related to shell oxidation and to partial, or total oxidation of the generated NPs. Here, it is reasonable to assume that only shell, or partial oxidation occurred, due to the dominant SPR peak, related to pure Cu. In the Ag NPs spectra [figure 5(c)] the extinction of the UV peak is considerably higher in (*iv*) for 10 J/cm², than in (*v*) for 20 J/cm², while the SPR peaks appear in both of them at about 408 nm, being most intense and narrowest in the latter.

Essentially, the single broad absorption bands from the UV to the Vis region are due to excitation of plasma resonances or interband transitions, and can be attributed to the presence of metallic NPs [22]. The distinct SPR absorption bands of Cu and Ag NPs in the Vis [21-22], could be affected by their sizes and shapes. Specifically, band positions and intensities were associated to the agglomeration extent , as well as to the size distribution of Ag NPs [26, 60-65] and to the shell, or partial oxidation of Cu NPs [24, 58, 66-69]. Since for Ag NPs very slight shifts in wavelength peaks positions *vs.* ablation fluences were observed, it is reasonable to assume that the mean diameter of the resulting particles was not changed for different fluences [61, 63].

To investigate the morphology and size distributions of the NPs and to reveal whether agglomeration plays a role under the ablation conditions of our experiments, as conjectured from the UV-Vis spectra, TEM measurements and analyses were conducted for most produced colloidal samples. Here, only the TEM images at high (A) and low magnification (B) of the samples of (a) Pd, (b) Cu and (c) Ag NPs obtained at laser fluences of 20 J/cm² [figure 6] and 10 J/cm² [figure 7] are shown. As expected from the UV-Vis spectra, these TEM images emphasize the principal changes in the prepared NPs. The generated NPs include very small particles, with diameters of less than 10 nm for all the three metals. Yet, some of the particles are much larger and spherical, probably due to NPs melting, required for NPs growing [29]. It is reasonable to assume that the melting is related to reinteraction of the NPs with the following laser pulses, if it is considered that the tails of the Gaussian laser beams have relatively low fluences. These fluences probably match the calculated melting fluences of Pd, Cu and Ag NPs with diameter < 20 nm, at a wavelength of 532 nm [29]. In addition, the larger NPs could probably be obtained through molten particles solidification, under the plasma plume.

The general appearances of the NPs, prepared with 20 and 10 J/cm² is completely different, while in the former they are less dense and only slightly agglomerated, it turns out that for the latter, clustering of the order of several μm becomes significant. In general, the NPs agglomerates, prepared at the additional laser fluences (1, 2, 5 and 90 J/cm²) were more prominent than at 20 J/cm². Considering these findings and the changes that were observed in the UV-Vis extinction spectra, figure 5, correlations between the agglomeration and the spectral behavior of the NPs of each metal could be suggested. Actually, no correlation between the largest agglomeration of the Pd, Cu and Ag NPs, obtained at fluence of 10 J/cm², and the highest concentrations at 5, 2 and 5 J/cm² (figure 3), respectively, was found. This finding attributed the agglomeration to the particle features, rather than to the high concentrations and strengthen the interpretation of the UV-Vis spectra, which distinct between the characteristic features of the particles at different fluences.

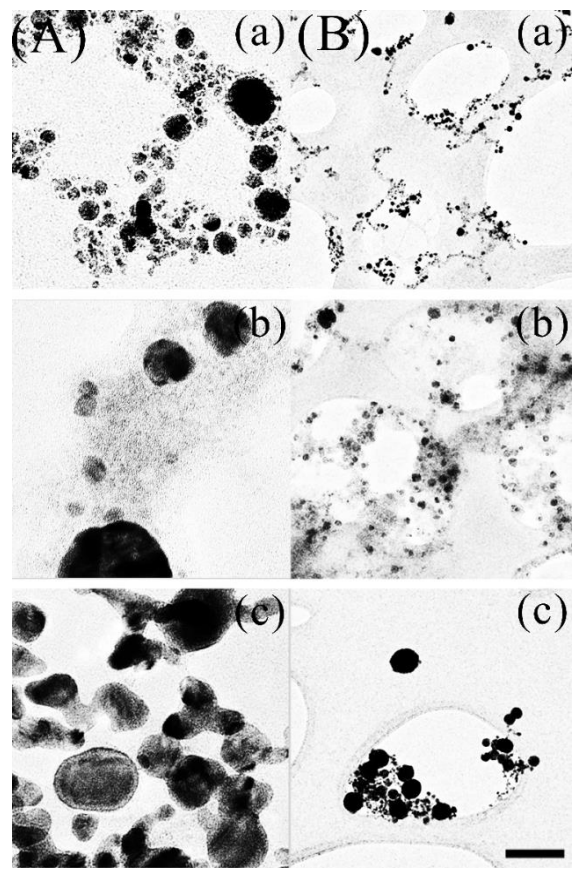

Fig. 6. Transmission electron microscope (TEM) images of (a) Pd, (b) Cu and (c) Ag nanoparticles (NPs), obtained by ablation at the interface of the meniscus of ethanol with the corresponding tilted metal targets, using a 20 J/cm² beam fluence. The images in (A) and (B) were taken at different magnifications and the scale bar at the right bottom represents 20 and 200 nm, respectively.

As mentioned above, the different agglomeration levels could be caused by the shape, size, defects, shell, or partial oxidation and other properties depending on unique laser-NPs interactions for the different fluences. It is conjectured that here the size of the NPs could be the reason for the agglomeration. For example, the high agglomeration of Cu [figure 7(A)(b)] corresponds to broadening and shifting of the SPR peak of the ablated Cu NPs, at a beam fluence of 10 J/cm² [figure 5(b)(*iv*)] and is associated with particles of less than 5 nm diameter [figure 7(A)(b)]. Nevertheless, the agglomeration process and the factors contributing to it, require further investigation.

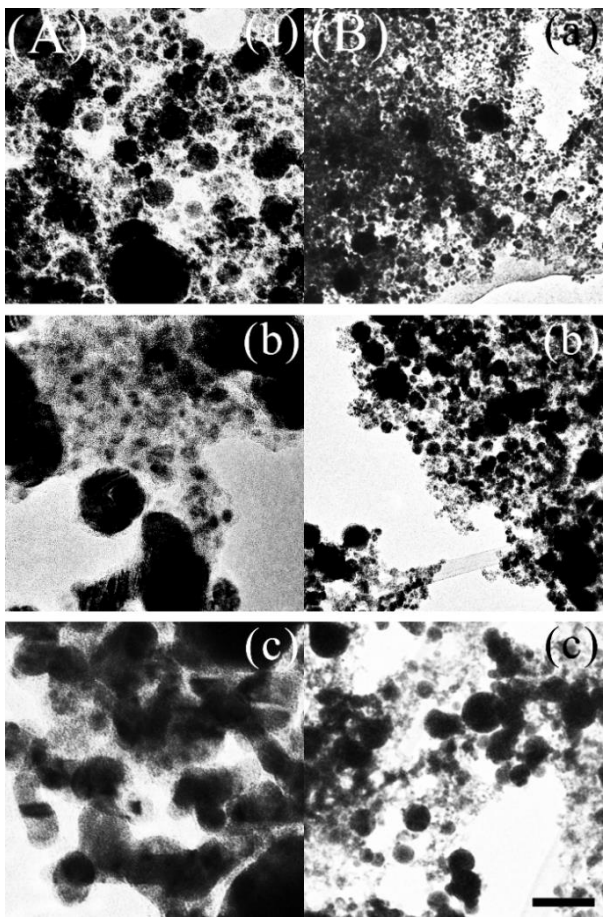

Fig. 7. Transmission electron microscope images of (a) Pd, (b) Cu and (c) Ag nanoparticles, obtained by ablation at the interface of the meniscus of ethanol with the corresponding tilted metal targets, using a 10 J/cm² beam fluence. The images in (A) and (B) were taken at different magnifications and the scale bar at the right bottom represents 20 and 200 nm, respectively.

Taking advantage of the high quantity of NPs that was prepared by the ablation at the interface of the meniscus of ethanol with the tilted metal targets, X-ray diffraction (XRD) spectra of (a) Pd, (b) Cu and (c) Ag particles were monitored. The typical spectra of the as-synthesized NPs are shown in figure 8. Each of the spectra includes characteristic diffraction peaks for Pd at $2\theta = 39.381°$, $45.832°$ and $66.821°$, corresponding to Miller indices (111), (200), (220), for Cu at $2\theta = 43.307°$, $50.476°$ and $74.229°$ assigned to the (111), (200) and (220) crystallographic planes and for Ag at $2\theta = 38.111°$, $44.294°$, $64.448°$ and $77.383°$ attributed to (111), (200), (220) and (311) planes. All the diffraction peaks of the metallic NPs correspond to face centered cubic (fcc) phases of crystalline Pd (ICDD card no. 01-087-0641), Cu (ICDD card no. 00-004-0836) and Ag (ICDD card no. 01-071-4613). This confirmed that the resultant Pd, Cu and Ag particles are with pure fcc structure.

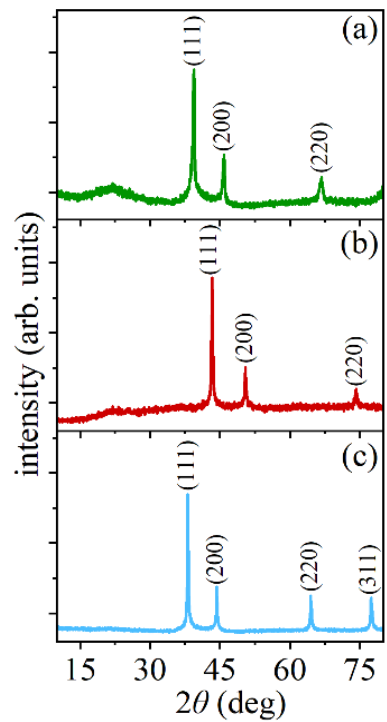

Fig. 8. X-ray diffraction patterns of (a) Pd, (b) Cu and (c) Ag NPs.

## 4. Summary

This study aimed toward optimization of the production of NPs and obtaining insight about their characteristics. Indeed, the use of the newly proposed PLAL configuration, where the Nd:YAG laser beam at 532 nm was directed toward interfaces of the menisci of ethanol with the tilted metal targets, facilitated Pd, Cu, and Ag NPs preparation. As foreseen and estimated from the calculated meniscus shape, this approach opened the possibility to reach the targets, by the laser beam, through thin ethanol layers, depending on ablation fluences. Moreover, the absorption in the colloidal solution was spatially bypassed by the falling of the NPs. The key result is that this unique and simple configuration resulted in improved productivities of small NPs, above those obtained by other methods [6], which could be represented by the power-specific productivity curves, peaking at particular fluences for the NPs of each monometal. Their shapes could be fitted, in the low fluence regime, to the results of a simple model accounting for $F_{eff}$, related to increased, or decreased laser energy deposited in plasma and vice versa in the target. Another uniqe advantage of these experiments is the extremly high concentrations of the NPs suspension. The resulting NPs were characterized by UV-Vis spectroscopy and TEM techniques, revealing the dependence of their extinction spectra and the extent of their agglomeration on the used laser fluence. In addition, the XRD technique allowed to reveal the NPs crystallinity. It is anticipated that this suggested strategy could be exploited in future studies for maximizing the yield of generated NPs and for revealing their properties. The spatial bypass of the cavitation bubble by the laser pulse that can be obtained through the use of this configuration, can assist in using higher repetition rate lasers and in enhancing the ablation efficiency.

## Acknowledgements

This work was supported by the Pazy foundation, under grant No. 286/18 and the Israel Science Foundation founded by The Israel Academy of Science.

# Supplementary Information for

# A simple strategy for enhanced production of nanoparticles by laser ablation in liquids: Experiment, modelling and characterization

Yaakov Monsa,[a] Gyora Gal,[b] Nadav Lerner[c] and Ilana Bar[a,*]

[a] *Department of Physics, Ben-Gurion University of the Negev, Beer-Sheva 8410501, Israel.*
E-mail

addresses:

*E-mail address*: ibar@bgu.ac.il
[b] *Department of Chemistry, Nuclear Research Center Negev, P.O. Box 9001, Beer-Sheva 8419001, Israel.*
[c] *Department of Analytical Chemistry, Nuclear Research Center Negev, P.O. Box 9001, Beer-Sheva 8419001, Israel*

### Menisci shapes

The calculated menisci shapes for a tilted Cu bulk target (2.10 rad) in ethanol (red), water (black) and acetone (green) are displayed in Fig. S1. These calculations were performed using the numerical approach of Ref. 1, while implementing it in a Mathematica code, using the NDSolve function and the given parameters in Table S1.

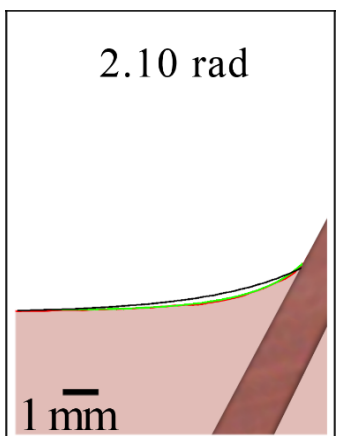

**Fig. S1.** Calculated menisci shapes, marked by lines, in ethanol (red), water (black) and acetone (green) with tilted bulk Cu metal targets at $\beta = 2.1$ rad (see Fig. 1).

**Table S1**. The parameters used for calculation of the menisci shapes of the different liquids with Cu targets.

|  | $\rho$ (kg/m³) at 25 ºC [i] | $\gamma$ (mN/m) at 25 ºC [i] | Outflow contact angle $\alpha$ (rad) for Cu |
|---|---|---|---|
| ethanol | 790.0 | 21.78 | 0.12 [ii] |
| acetone | 781.8 | 22.67 | 0.25 [iii] |
| water | 997.1 | 71.87 | 0.58 [ii] |

[i] Ref. 2
[ii] Ref. 3
[iii] Ref. 4

The same procedure was also used for calculation of the menisci shapes of ethanol with bulk Cu targets, set at different inclination angles, and the results are shown in Fig. S2.

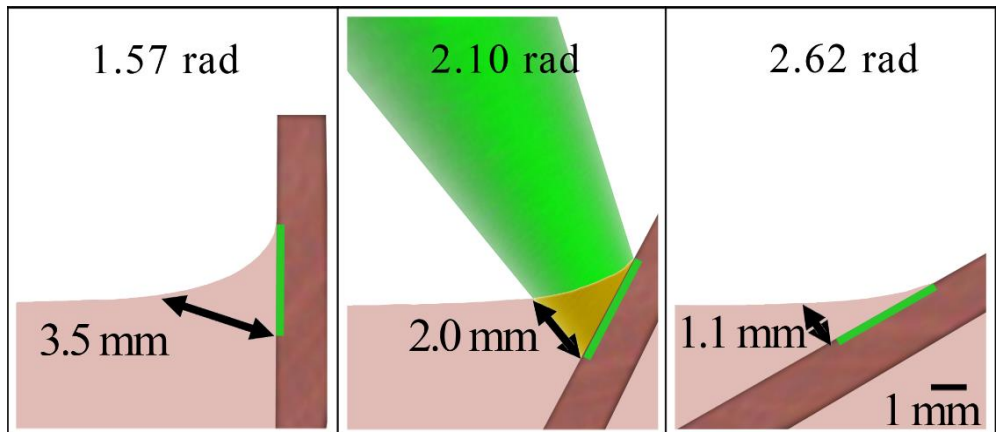

**Fig. S2.** Calculated menisci shapes of ethanol with tilted bulk Cu metal targets, for different $\beta$ values (see Fig. 1). The beam and the green lines depict the laser spot diameter accessing the target at the lowest used fluence of 0.6 J/cm² and the black arrows and the values in mm mark the sampled maximal thicknesses of ethanol above the targets.

**UV-Vis absorbance spectra**

The originally measured UV-Vis absorbance spectra for Pd, Cu and the Ag NP, prepared with different laser fluences are shown in Fig. S3. These spectra are similar to those shown in Fig. 4, which have been vertically translated for clarity.

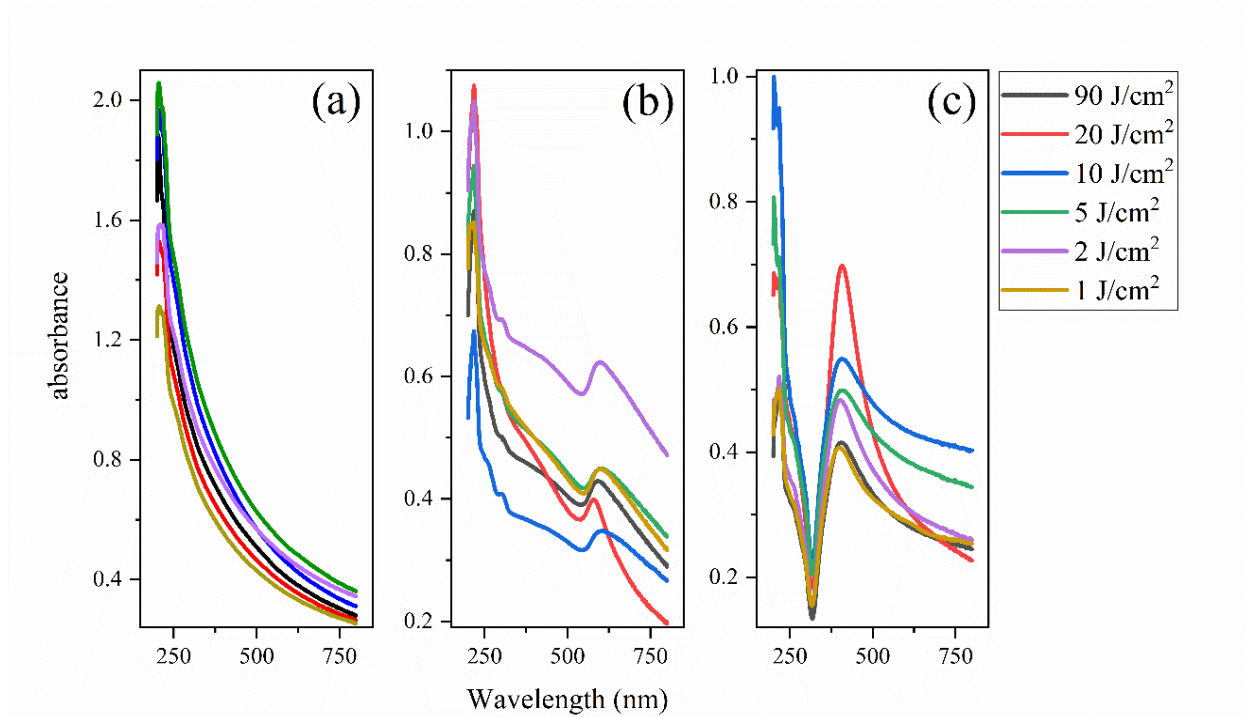

**Figure S3.** Measured UV-Vis absorbance spectra of diluted suspensions of (a) Pd, (b) Cu and (c) Ag NPs in ethanol, shown in different colors, representing the laser fluences that were used for their preparation and given in the upper right corner.